\documentclass[iop,apj]{emulateapj}
\usepackage{amsmath, amsthm, amssymb, amsfonts}
\usepackage{graphicx}
\usepackage{dcolumn}
\usepackage{hyperref}
\usepackage{natbib}
\usepackage{bm}
\usepackage[caption=false]{subfig}
\usepackage{verbatim}

\begin{document}

\title{The Polarization Behavior of Relativistic Synchrotron Self-Compton Jets}

\author{A. L. Peirson\altaffilmark{1} \& Roger W. Romani\altaffilmark{1}}
\affil{\altaffilmark{1}Dept. of Physics and Kavli Institute for Particle Astrophysics and Cosmology, Stanford University, Stanford, CA 94305}
    
\begin{abstract}
We describe a geometric model for synchrotron and synchrotron self-Compton (SSC) radiation from blazar jets, involving multiple emission zones with turbulent magnetic fields and fully self-consistent seed photon mixing for SSC. Including the effects of jet divergence, particle cooling and the Relativistic PA rotation (RPAR) to the observer frame, we find that the multi-zone model recovers simple predictions for SSC polarization, but describes new dependencies on jet viewing geometry and zone multiplicity. Increasing the zone number decreases both synchrotron and SSC polarization, but with different scaling. A rise in synchrotron polarization fraction $\Pi_{\rm Sync}$ at high energies is guaranteed by basic relativity considerations, and strengthened by jet non-uniformity. Finite light travel time effects can suppress the synchrotron polarization at energies well below the $\nu_{\rm Sync}$ peak. In general $\Pi_{\rm Sync}$ and $\Pi_{\rm SSC}$ are correlated with $\Pi_{\rm SSC}/\Pi_{\rm Sync}\approx 0.3$, but individual realizations can lie far from this trend. This study lets us estimate $\Pi$ across the SED, leading to predictions in the X-ray band helpful for planning observations with {\it IXPE} and other upcoming X-ray polarization missions.
\end{abstract}

\pacs{Valid PACS appear here}
\maketitle


\section{\label{sec:level1}Introduction}
Blazars are active galactic nuclei whose powerful relativistic jets point at small angle $\theta_{\rm obs}$  to the Earth line-of-sight \citep{urry_unified_1995}, so that the Doppler-boosted jet emission dominates the observed spectral energy distribution (SED). This SED is characterized by a low energy peak caused by synchrotron radiation from energetic electrons, and a high energy peak generally attributed to inverse-Compton (IC) scattering of photons by these same electrons \citep{maraschi_jet_1992}, a.k.a. synchrotron self-Compton (SSC). The seed photons can also originate from an external source such as the accretion disk or broad line region (External Compton, EC). Blazars can be subdivided by the frequency of their synchrotron peak \citep{abdo_spectral_2010} into HSP, LSP and ISP sources. HSP tend to to peak in the X-ray.
We have yet to determine how the jets are energized and launched with bulk Lorentz factor $\Gamma$, but an attractive origin is the \citet{blandford_electromagnetic_1977} process, so that the jet axis may be associated with the spin axis of the central black hole and the angular momentum axis of the surrounding accretion disk. The jet $e^+/e^-$ obtain an energy distribution extending to $\gamma_{\rm max} \sim 10^4$ or higher, often attributed to shock acceleration or magnetic reconnection. Radiation from these particles spiraling in the embedded magnetic field can be used to constrain the geometry and energetics of the emission zone and, by inference, the jet accelerator.

In studying jet geometry polarization can be particularly useful. Radio VLBI studies have long shown that the pc-scale jet can be substantially polarized. Recently much effort has been spent measuring the optical polarization properties of blazars, since this probes even smaller scales, closer to the acceleration zone. This polarization is often quite variable, offering new dynamical information on the jet structure \citep[e.g.][]{blinov_robopol:_2015, lynch_green_2018}.
Soon we hope to measure the X-ray polarization of a number of blazars with \textit{IXPE} \citep{weisskopf_imaging_2016}, probing closer to the jet acceleration zone than ever before.

Recent optical monitoring campaigns have revealed new polarization patterns. In addition to typical stochastic behavior of polarization fraction ($\Pi$) and angle (PA), \citet{blinov_robopol:_2015} found periods of relatively steady rotation of the PA, sometimes extending many $\times \pi$, lasting weeks or months. These rotations are sometimes associated with flares in total intensity and drops in $\Pi$ \citep{blinov_robopol:_2016}. Various models have been proposed to explain this behavior, including a turbulent stochastic model \citep{marscher_turbulent_2014}, a spiraling jet \citep{lyutikov_polarization_2017} and a helical kink propagating along a conical jet \citep{nalewajko_model_2017}. Although these pictures can accommodate multicycle rotations, they fail to address the optical trends found in \citet{blinov_robopol:_2016}.

In \citet{peirson_polarization_2018}, we modeled a simple multizone conical jet model, optionally with a helical core field, and found that, when a proper treatment of relativistic PA rotation (RPAR) is included, it can explain many of the synchrotron emission trends mentioned above.
In addition it makes a number new, testable predictions, which can help in interpreting future optical/IR polarization campaigns. However, we soon expect to measure X-ray polarizations with space missions such as {\it IXPE}; This band is often well above the $\nu_{Sync}$ synchrotron peak and, for LSP and ISP, may include significant Compton flux. 

In this paper we extend our conical jet model to include the multizone treatment of SSC polarization in blazars. We particularly focus on the transition region between synchrotron and SSC dominated flux, as this will be {\it IXPE}'s range for many ISP. We start by reviewing the residual polarization after Compton scattering. We then describe how the synchrotron emission of multiple zones are combined for Compton re-processing. \S4 notes how light-travel time effects additionally modify the polarization seed flux; this varies along the jet so that the effects vary with energy band. A numerical realization of this model lets us check how the averaging scales with the number of effective jet zones (which can be constrained by e.g. optical $\Pi_O$).  We conclude with full Sync+SSC simulations for representative blazar parameters.

\section{\label{sec:level1}ICS Basics \& Seed Photon Polarization}

\citet{bonometto_polarization_1970} have developed an analytical formalism to evaluate inverse-Compton polarization for scattering in the Thomson scattering regime, with \citet[][hereafter BCS]{bonometto_polarization_1973} treating the SSC case. Analytic solutions are difficult in the Klein-Nishina regime, but \citet{krawczynski_polarization_2011} provides a general Monte-Carlo based framework, verifying and extending the BCS results. To date such pseudo-analytic treatments have been applied to homogeneous single-zone jet models. The results provide useful upper limits on the plausible ICS polarization. For example \citet{poutanen_relativistic_1994} explore how the magnetic field orientation in a uniform jet affects $\Pi_{SSC}$, while \citet{mcnamara_x-ray_2009} run a single zone Monte-Carlo model to show how X-ray polarization would differ between synchrotron, SSC and Externally-dominated Compton (EC) emission. Finally \citet{zhang_x-ray_2013} argued from single zone SSC simulations that polarization measurements can distinguish between weakly polarized leptonic Compton emission and strongly polarized hadronic models. 

However, with the new (especially optical) evidence for incoherent jets with multiple zones contributing to the polarized flux, these studies are inadequate to describe any but the most basic differences between jet models. In addition, many of these previous efforts do not fully incorporate the important RPAR rotation of the emitted polarization to the Compton scattering zone and on to the observer frame. \citet{marscher_turbulent_2014} introduce a multizone framework; in \citet{peirson_polarization_2018} we explored synchrotron emission in this picture, including the effects above. Here we summarize the jet geometry before describing additional polarization averaging of the seed synchrotron flux and computing the final Compton polarization.

Our jet is conical with opening angle $\theta_{op}$ pointing $\theta_{obs}$ from our line of sight. The cross sectional radius at launch $R_0$ is determined by the jet power $W_j$, initial magnetic field strength $B_0$ and bulk Lorentz factor $\Gamma$ (all values set in the jet frame) assuming an equipartition fraction $= 1$. The jet is segmented into slices (sections) along the jet, each made up of multiple zones $i$, which share the same bulk $\Gamma$ but have different $\theta_{obs_i}$. The B-field orientation varies, typically randomly, between zones. Alternatively a subset (assumed to be the jet core) has a coherent helical B-field during polarization rotation epochs. 
Each zone has an initial electron population set by power law index $\alpha$ and exponential cutoff set by $\gamma_{\rm max}$.
As a given slice moves down the jet, at each step $dx$ the polarized synchrotron emission from each zone is calculated using expressions from \citet{rybicki_radiative_1979}. $B$, $R$ and the electron populations are evolved at each step. Applying relativistic PA rotation (RPAR) to the emission of each zone and summing the Stokes parameters gives the final SED and polarization. 

To focus on overall geometrical trends our base calculations assume that all zones are identical except for field orientation. This allows us to ignore electron migration between zones in a given slice; we also ignore small losses associated with complete escape from the jet. Variation in zone efficiency should introduce additional variability, diluting but preserving the geometric trends described here. We do discuss (\textsection3,5) cases when a subset of zones dominate the synchrotron emission, since the ICS emission can be sensitive to their disposition across the jet.



\begin{figure}[]
\includegraphics[width=0.9\linewidth, height=7.0cm]{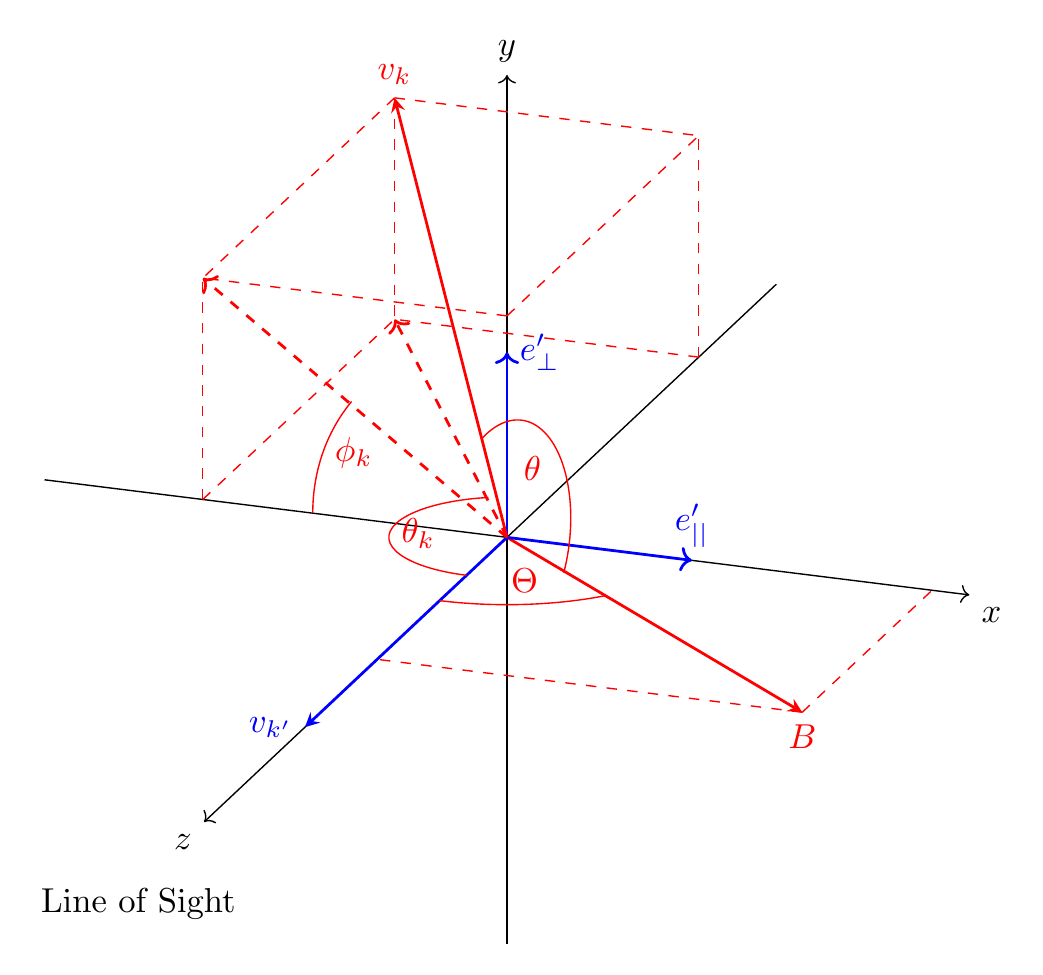}
\centering
\caption{Sketch of the SSC scattering geometry to illustrate the definition of angles. The coordinate system is chosen such that the magnetic field B lies in the (x-z) plane and the scattered photon direction ${\bf v_{k'}}$ is along the z axis.}
\end{figure}

As we are most interested in X-ray and lower energies, we treat SSC in the Thomson limit, which \cite{zhang_x-ray_2013} show is valid up to at least $500$MeV for a relativistic jet with Doppler factor $D \gtrapprox 10$. The incoming and outgoing photons have momentum unit vectors $\bf{v_k},\bf{~v_{k'}}$ and frequencies $\epsilon$, $\epsilon'$ respectively. BCS show that photons with original polarization direction $\bf{e}$ (perpendicular to the magnetic field for synchrotron seed photons) scatter to energy $\epsilon'$ with powers $P^{SSC}_{||}$ and $P^{SSC}_{\perp}$ for polarization along or perpendicular to the projection of the B-field onto the plane orthogonal to $\bf{v_{k'}}$:
\begin{equation}
    P^{\rm SSC}_{||}(\epsilon') = C\epsilon'\int\frac{d\epsilon}{\epsilon}d\Omega_kE_{\rm min}n(\epsilon)q(\theta) \cdot (Z_{||}(\Sigma_1 + \Sigma_2) + \Sigma_2)
\end{equation}
\begin{equation}
    P^{\rm SSC}_{\perp}(\epsilon') = C\epsilon'\int\frac{d\epsilon}{\epsilon}d\Omega_kE_{\rm min}n(\epsilon)q(\theta) \cdot (Z_{\perp}(\Sigma_1 + \Sigma_2) + \Sigma_2).
\end{equation}
Here $C = \pi (\frac{e^2}{4\pi})^2\frac{c}{m_ec^2}$ in c.g.s units,
\begin{equation}
    E_{\rm min} = \sqrt{\frac{\epsilon'}{2\epsilon(1-{\rm cos}\theta_k)}}
\end{equation}
is the minimum electron energy required for scattering of a photon from $\epsilon$ to $\epsilon'$,
\begin{equation}
    Z_{\bf{e'}} = \bigg(\mathbf{e \cdot e'} + \frac{(\mathbf{v_k \cdot e'})(\mathbf{v_{k'} \cdot e})}{1-{\rm cos}\theta_k}\bigg)^2
\end{equation}
where $Z_{||}$ and $Z_{\perp}$ are found by selecting $\mathbf{e'}_{||}$ or $\mathbf{e'}_{\perp}$ respectively (fig.1), and the solid angle of the photon direction before scattering is
\begin{equation}
    d\Omega_{k} = d{\rm cos}\theta_kd\phi_k.
\end{equation}
$\Sigma_1$ and $\Sigma_2$ are integrals over the electron population doing the scattering, with maximum electron energy $E_2$ and minimum $E_1$:
\begin{equation}
    \Sigma_1 = \int^{\beta_1}_{\beta_2}dE\frac{N_e(E)}{E^4}\bigg(\frac{E_{\rm min}^2}{E^2} - \frac{E^2}{E_{\rm min}^2} + 2\bigg)
\end{equation}
\begin{equation}
    \Sigma_2 = \int^{\beta_1}_{\beta_2}dE \frac{N_e(E)}{E_{\rm min}E^6}(E^2 - E_{\rm min}^2)^2
\end{equation}
where $E$ is the electron energy and
\begin{equation} 
\beta_1 =
  \begin{cases}
    E_{\rm min}       & \quad  E_{\rm min} > E_2\\
    E_2  & \quad  E_{\rm min} < E_2\\
  \end{cases}
\end{equation}
\begin{equation}
\beta_2 =
  \begin{cases}
    E_{\rm min}       & \quad  E_{\rm min} > E_1\\
    E_1  & \quad  E_{\rm min} < E_1\\
  \end{cases}
\end{equation}
$n(\epsilon)$ and $q(\theta)$ denote the synchrotron seed photon spectrum, split into an energy and angle dependent part, where the angle $\theta$ is given by 
\begin{equation}
    {\rm cos}\theta = {\rm cos}\Theta {\rm cos}\theta_k + {\rm sin} \Theta {\rm sin} \theta_k {\rm cos} \phi_k 
\end{equation}
from fig. 1. We take $q(\theta) \propto {\rm sin}^{\frac{p+1}{2}}\theta$, given an isotropic distribution of electron pitch angles. $n(\epsilon)$ is calculated self-consistently from the multizone model at each step of the jet. 

 The framework described above assumes a 100\% polarized seed photon population. BCS treated only power law electron populations with synchrotron polarization independent of $\epsilon$ (i.e. $\Pi(\epsilon) =$ const.). For partly polarized seed photons \citet{bonometto_polarization_1970} ignored energy dependence, simply re-scaling the final SSC polarization fraction $(P^{\rm SSC}_{\perp} - P^{\rm SSC}_{||})/(P^{\rm SSC}_{\perp} + P^{\rm SSC}_{||})$. In our case the electron population cools, so that $E_c$ and the photon spectrum evolve, meaning that we cannot assume constant $\Pi(\epsilon)$. Thus we split $n(\epsilon)$ into $n_{\perp}(\epsilon)$ and $n_{||}(\epsilon)$ (synchrotron photon populations with polarization parallel and perpendicular to the projection of the B-field in the plane orthogonal to $\bf{v_k}$) where $\Pi(\epsilon) = (n_{\perp}(\epsilon) - n_{||}(\epsilon))/(n_{\perp}(\epsilon) + n_{||}(\epsilon))$. Evaluating these separately using (1) and (2), we sum their Stokes' parameters to get the SSC polarization for arbitrary $\Pi(\epsilon)$. 
 
With jets having significant bulk $\Gamma$, we expect blazar emission to be affected by RPAR (\citealp{peirson_polarization_2018, lyutikov_polarization_2003}). This relativistic aberration strongly changes our effective line of sight, thus rotating the PA we observe as a function of $\Gamma$ for fixed $\theta_{obs}$. We have shown RPAR to be relevant in both stochastic and rotation phases in blazar synchrotron polarization \citep{peirson_polarization_2018}; we expect it to be even more important here since SSC polarization is strongly dependent on the component of the B-field to our line of sight \citep{bonometto_polarization_1973}; note the $\Theta$ dependence in Eqn. 10 and the $\mathbf{e}$ vector.
We include the effects of RPAR in our model by rotating the jet frame $\mathbf{B'}$ to the effective magnetic field observed in the lab frame for each zone 
when calculating the SSC emission and final Stokes' parameters.
\begin{equation}
    \mathbf{B_{eff}} = \mathcal{R}(\Theta_\mathrm{rot}) \cdot \mathbf{B'}
\end{equation}
and
\begin{equation}
    \Theta_\mathrm{rot} = \arccos\bigg(\frac{\cos\theta_{vl}-\beta}{1-\beta \cos\theta_{vl}}\bigg) - \theta_{vl}~,
\end{equation}
where $\theta_{vl}$ is the angle between the zone's velocity vector $\beta c \hat{\mathbf{v}}$ and our line of sight in the lab frame. The rotation takes place along the plane containing $\beta c \hat{\mathbf{v}}$ and the line of sight in the lab frame. This is a simpler more intuitive form of the RPAR equations given in  \citealp{peirson_polarization_2018, lyutikov_polarization_2003}.

This prescription gives us the observed synchrotron and SSC polarized emission from a single B-field zone, assuming that all Compton upscatter is only from local sychrotron seed emission (`on the spot' approximation). This is, of course, the approximation used in single zone models. Instead we expect that the true seed photon field will be strongly dependent on the inhomogeneous surrounding zones \textsection3. Further, since the jet electron populations evolve, the seed photons seen at a given zone are also dependent on light travel effects, which we discuss in \textsection4. In particular, since the low energy seeds dominating the upscattering tend to be dominated by the cooled population, this can be especially important for low energy (e.g. X-ray) SSC emission. Nevertheless for some initial insight, we start by computing emission from a multi-zone jet, with SSC independently computed for each zone, as above.

We proceed by computing the SSC Stokes components for each B-field zone individually, evolving the electron population and jet parameters by calculating the total electron energy losses at each $dx$ step, then summing the Stokes' flux across the full length of the jet. In this evolving, but isolated, zone example the energy density depends only on $R$ and the instantaneous emitted synchrotron power, and so is the same at all points in a given jet cross-section. Figure 2 shows a simulation slice with isolated zones and typical blazar parameters for synchrotron + SSC. Note the sharp rise in $\Pi_{Sync}$ and $\Pi_{SSC}$ (and EVPA shift) at the upper end of each component. This is more fully explored in \textsection3.

\begin{figure}[]
\includegraphics[width=0.98\linewidth, height=7.0cm]{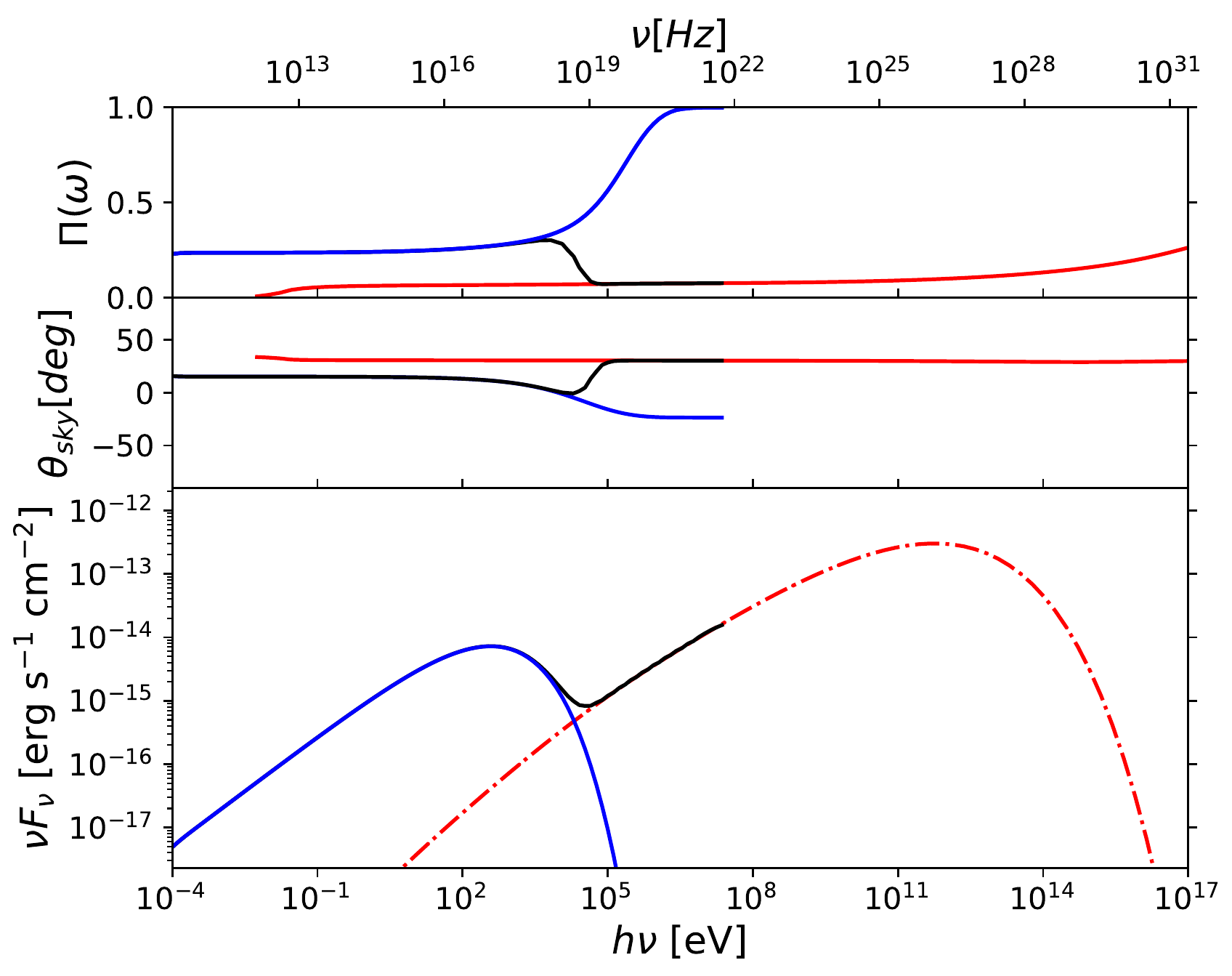}
\centering
\caption{The polarization fraction, observed PA angle and SED for a jet with 19 independent B-field zones jet for a single $dx$ step (i.e. no cooling). The SSC emission is calculated using only local synchrotron seed photons (i.e. evaluating Eq. (17) instantaneously). Published models typically use such a single zone. Red denotes SSC, blue synchrotron and black their combination. The jet parameters are tabulated in the Appendix.
}
\end{figure}

Before we extend to the interacting zone model, we mention some general results already visible in these sums. First, as noted by \citet{bonometto_polarization_1970}, unlike Thomson scattering, Compton scattering does not create polarization. Thus inevitably $\Pi_{\rm SSC} < \Pi_{\rm Sync}$ (although for multi-zone seed photon mixing, this is not always true, \textsection5). Next, a $\Pi_{\rm Sync}=1$ beam scattering off an $e^-$ powerlaw of index $\alpha$ will produce $\Pi_{SSC} = \Sigma_1 + \Sigma_2 / (\Sigma_1 + 3\Sigma_2)$, with the EVPA reflected in the $\mathbf{v_{\rm k}}$ and $\mathbf{v_{\rm k'}}$ plane. For typical values of $\alpha \sim 1-3$, $\Pi_{\rm SSC} \sim 0.5-0.75$. The modest $\Pi_{\rm Sync} \approx 0.03-0.1$ of real jets indicates many $N_{\rm eff}$ emission zones with uncorrelated B field orientations. While $\Pi_{\rm SSC}$ will depend on the particular B orientations of a given realization, as $N_{\rm eff}$ increases, the result tends to an isotropic average. For a single isolated zone averaged over many isotropic B-field realizations we find $\frac{\Pi_{\rm SSC}}{\Pi_{\rm Sync}} \approx 0.35$, 
(in good agreement with the \citet{bonometto_polarization_1973} result for $\alpha \sim 2$). Note that $\Pi_{\rm Sync}$ is the polarization of the typical seed photons (e.g. $0.01-0.1$eV for for X-ray SSC), discussed in \textsection 5.

\begin{figure}[]
\includegraphics[width=0.96\linewidth, height=7.3cm]{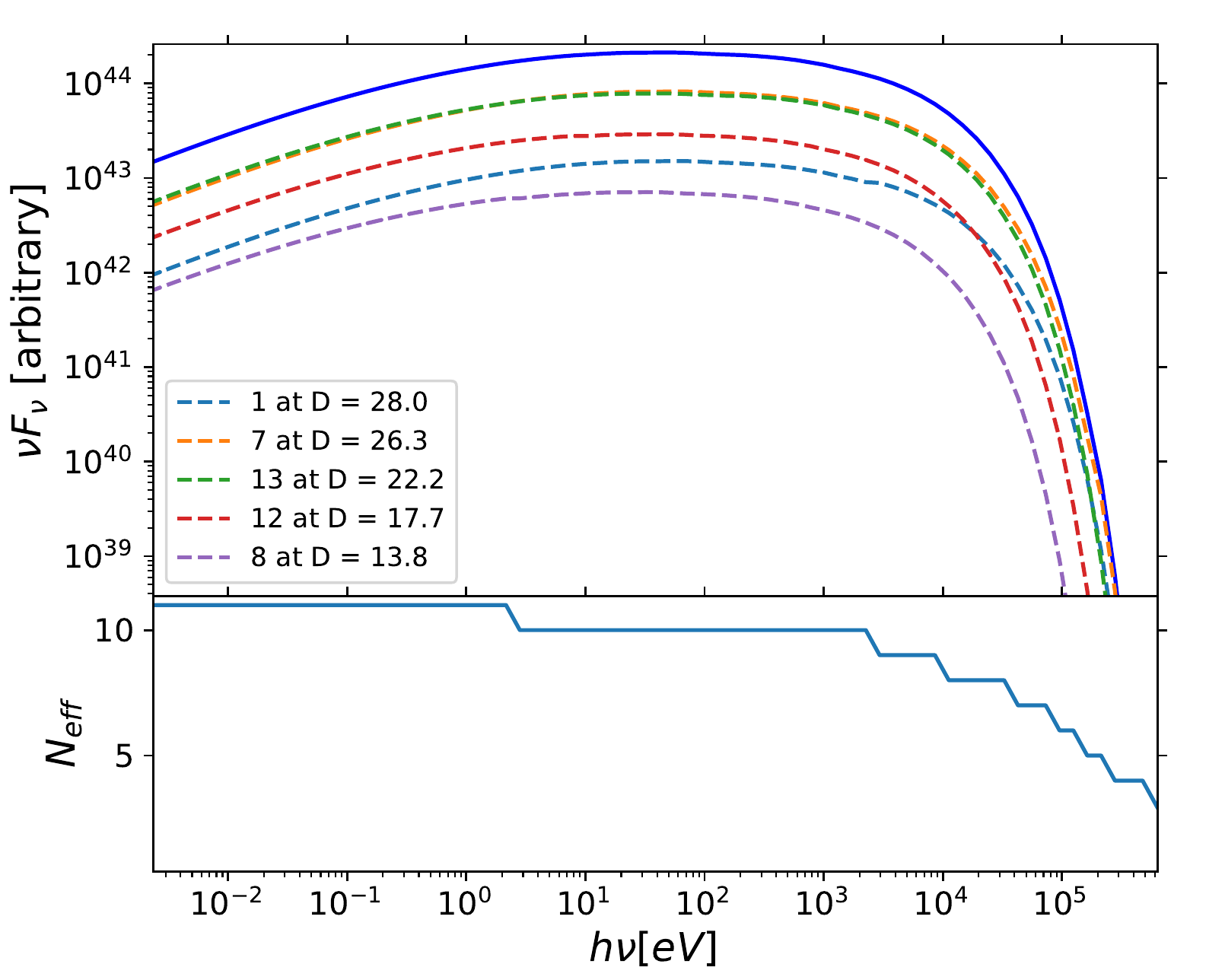}
\centering
\caption{Plot of an example full jet synchrotron SED (blue solid) made up of emission from 19 B-field zones. The zones are grouped into 5 different Doppler factor sets. Dashed lines show the SEDs from zone sets with differing $\theta_{\rm obs_i}$. We can see that the total emission (blue solid) has contributions from fewer total zones at high energy. Once the SED is exponentially dominated, the zone with the highest $\gamma_{\rm max}$ dominates the observed flux.}
\end{figure}

\begin{figure}[]
\includegraphics[width=0.96\linewidth, height=7.3cm]{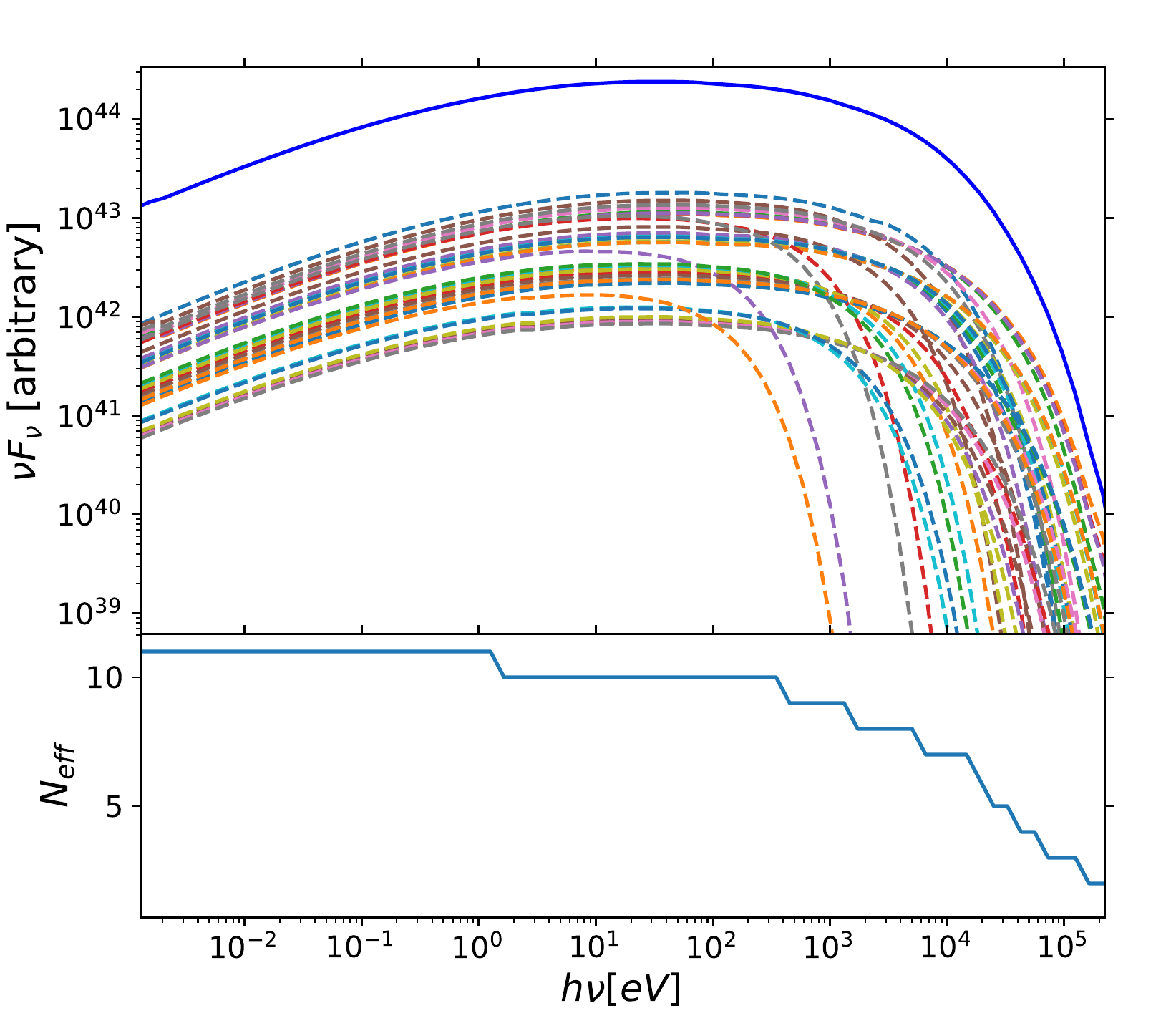}
\centering
\caption{Effect of combined Doppler boosting and $\gamma_{\rm max}$ spread at the synchrotron SED. Each dashed line now represents an individual zone.  Again, $N_{\rm eff}$ drops at high energy.}
\end{figure}

\section{\label{sec:level1}Multi-zone Effects}

In a conical $\theta_{\rm op}$, multizone, fixed-$\Gamma$ jet model, each B-field zone $i$ is observed at a different $\theta_{\rm obs_i}$ and thus has Doppler factor $D_i = \frac{1}{\Gamma(1-\beta {\rm cos}\theta_{\rm obs_i})}$. The observed (lab) power of each zone is then $F_{\rm lab}(\nu) = D_i^3 \cdot F_{\rm jet}(\nu)$ where two powers of $D_i$ come from relativistic aberration and the other from time dilation. Additionally, the frequencies in the lab frame are blue-shifted, $\nu_{\rm lab} = D_i\cdot \nu_{\rm jet}$. For a power law photon spectrum $F(\nu) \propto \nu^{-\beta}$, this provides an additional factor of $D_i^{-\beta}$ to the energy spectrum. 
For the Doppler boosting computations, we assume that the jet is structured as a set of discrete blobs, as observed at VLBI scale. 
Thus non-cylindrical geometry alone, with Doppler boosting, guarantees that identical jet zones contribute differently to the the observed synchrotron peak (Fig. 3) and to the seed photon population seen by other zones. 

Note that with increasing observed photon energy one samples further into the exponential tail of the individual zone spectra. Thus a decreasing number of zones contribute until a single zone dominates. The result is an increasing $\Pi_{\rm Sync}$ and a gradual EVPA evolution, converging on that of the most boosted zone (in the few-zone regime this behavior may not be monotonic). The top panel of figure 2 showcases this effect, while figure 3 shows an example of individual zone spectra. A closely related effect occurs when the zones themselves have different $\gamma_{\rm max}$. For example \citet{marscher_turbulent_2014} assumes in a jet-shock model that $\gamma_{\rm max}$ depends on B field-shock inclination angle, for an injected electron spectrum of index $\alpha$, giving
\begin{equation}
N(E_e) \propto E_e^{-\alpha}e^{\frac{-E_e}{\gamma_{\rm max}m_ec^2}}
\end{equation}
with
\begin{equation}
\gamma_{\rm max} \propto \bigg(\frac{B_{||}}{B} \bigg)^2 .
\end{equation}
and $B_{||}$ the zone's B-field component parallel to the shock normal. Thus this version is sensitive to the shock geometry. Alternatively we might imagine that shock turbulence gives rise to the same $\gamma_{\rm max}$ distribution as (14) but with orientation independent of the shock geometry, as for our fully random B distribution. In any case, the geometric Doppler effect combines with the intrinsic $\gamma_{\rm max}$ effect to disproportionately weight a subset of the zones. Figure 4 shows the zone spectra when both effects are present. Figure 5 gives the effect on net polarization, and its dependence  on jet parameters. Here we define $N_{\rm eff}$ as the number of zones contributing half of the integrated flux. Clearly when $\gamma_{\rm max}$ scales with the shock $B_{||}$ one finds the highest polarization fraction, since the dominating zones have B-fields nearly aligned (although we note that when the shocks are transverse, this large $\Pi$ is strongly dependent on RPAR effects). Interestingly for large $\theta_{\rm op}$, Doppler boosting alone can produce close to the same $\Pi$ rise as a $\gamma_{\rm max}$ spread.

Even without any zone differences, a small increase in $\Pi$ is expected at frequencies emitted by electrons above $\gamma_{\rm max}$ due to the deviation from a perfect power law. Thus $\Pi$ tends to 1, not $\sim 0.75$, when a single zone dominates on the exponential tail.

\begin{figure}[]
\includegraphics[width=1.0\linewidth, height=7.3cm]{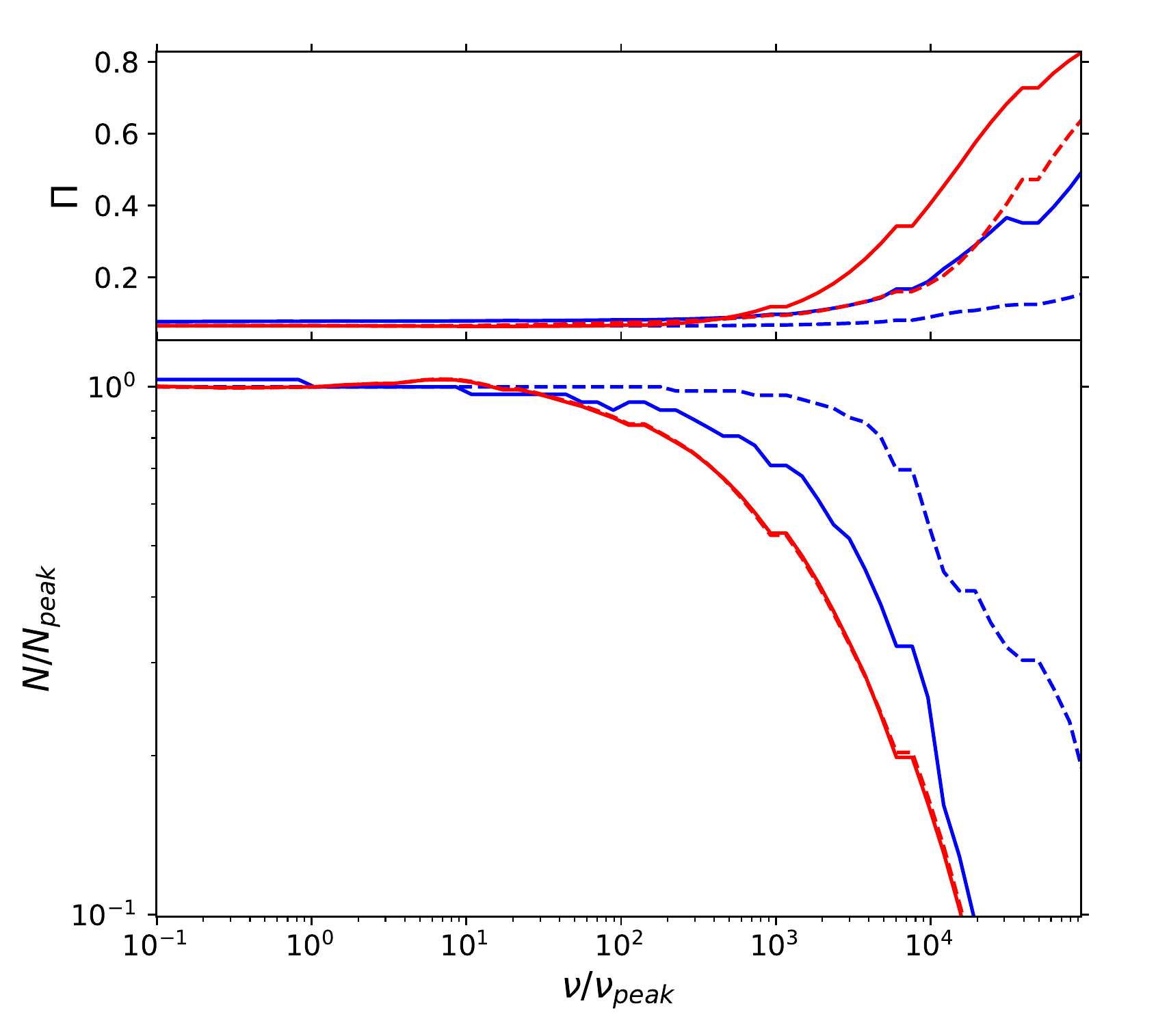}
\centering
\caption{\textit{Bottom Panel:} The number of zones contributing half of the observed flux (normalized to the number contributing at the  SED peak) for a typical full jet blazar synchrotron SED. The blue lines represent the dependence for the Doppler effect of Fig. 3. The solid line shows a jet with large $\theta_{\rm op}$ viewed near its edge (large Doppler spread) while the dashed line shows a small $\theta_{\rm op}$ viewed near the axis (small Doppler spread). Red curves give the case where $\gamma_{\rm max}$ varies between zones. The solid red curve is the scenario proposed in \cite{marscher_turbulent_2014} where $\gamma_{\rm max}$ depends on $B_{||}$, the shock-aligned field (here we assume a transverse shock). The dashed red curve represents the same $\gamma_{\rm max}$ distribution as the solid line case, but with random $B$ orientation. \textit{Top Panel:}  The corresponding polarization fractions; colors and lines as below.}
\end{figure}

\section{\label{sec:level1}Light Travel Effects}

Since our multizone jets have evolving (cooling) electron populations and since B may also be a function of time or distance along the jet, we need to consider how the finite light travel time (and the slower jet speed) affect the emission observed at any one moment. This is especially important when considering propagation between jet zones in building up the SSC seed photons. First, we should recall that our multizone model has a spatial coherence scale across the jet, the distance over which the magnetic field (and possibly $\gamma_{\rm max}$) decorrelate. With $N$ zones this is $\approx 2R_0/\sqrt{N}$ for the stochastic magnetic field. Assuming isotropic jet turbulence, this should also be $dx_{\rm ch}$, the decorrelation length {\it along} the jet in the jet frame -- this gives a decorrelation timescale $\sim dx_{\rm ch} / c$. For a range of blazar parameters we find $dx_{\rm ch} \sim 10^{15} - 10^{16}$cm giving polarization variability timescales of $\sim 0.5-5$\,d, in agreement with the stochastic optical variability measured by RoboPol \citep{liodakis_prospects_2019}.

For rotation-dominated epochs with the helical B fields we can instead associate $dx_{\rm ch}$ with the timescale of a typical observed optical rotation, roughly $360^{\circ}$ per month \citep{blinov_robopol:_2018}, so we take $dx_{\rm ch} \approx 1{\rm d} c/\beta$. For many typical blazar parameters these stochastic and helical characteristic length scales are of similar size. We consider how the observed spectrum and polarization are sensitive to this coherence scale.

\subsection{Finite Bulk Lorentz factor}

For infinite $\Gamma$, the jet particles and their emitted radiation would be co-spatial for their entire radiation history (and we would detect this flux only along the jet axis). However with finite $\Gamma$ the photons outrun the jet particles. If at some energy the dominant radiation is produced sufficiently far downstream it will lag behind the radiation produced closer to the jet base by that same zone. Thus at a distance $x_p = \beta dx_{\rm ch} / (1 - \beta)$ the radiation from our designated zone will not yet have reached the Earth observer; we will instead measure the flux of the {\it preceding} zone (along the same jet flow line). This preceding zone will in general have different $B$ and particle population properties. Further downstream additional zones can also contribute.

However in practice for $\Gamma > 5$ and $N \leq 150$ we find that the bulk of the observed emission at all frequencies of interest has been radiated before $x_p$ (as can be confirmed with the bottom panel of fig. 7). Thus we can infer that the radiation from a single slice is co-eval, except for the most extreme jet parameters. 

In this picture the field orientation at the jet base is frozen in and $dx_{ch}\sim$ const, so that zones expand only transversely. If in contrast the zones stay quasi-spherical (e.g. due to turbulent cascading along the jet), a longer variability timescale and a decorrelation in polarization compared to higher frequencies can result for late jet emission (radio). 

\subsection{Non-zero Viewing Angle of a Conical Jet}

For a diverging jet viewed off axis, the increasing width of an observer time slice includes an increasing range of jet distances (i.e. larger range of emission times for the jet particles). This is shown in Figure 6. In our zonal picture, this means that once $2R {\rm tan}\theta_{\rm obs} / \sqrt{N} > dx_{\rm ch}$ zones from more than one slice contribute to the emission. For a given $N$ zones in a jet, expansion stretches the zone horizontally, but not radially. Thus with a tilt, the increased radial range incorporates more $B$-field zones at a given observer time slice, and the polarization decreases. 
In practice, this is dependent on the jet opening angle $\theta_{\rm op}$ through both the expansion rate of $R$ and the individual zone Doppler factors that control $dx_{\rm ch}$.

\begin{figure}[]
\includegraphics[width=0.9\linewidth, height=5.3cm]{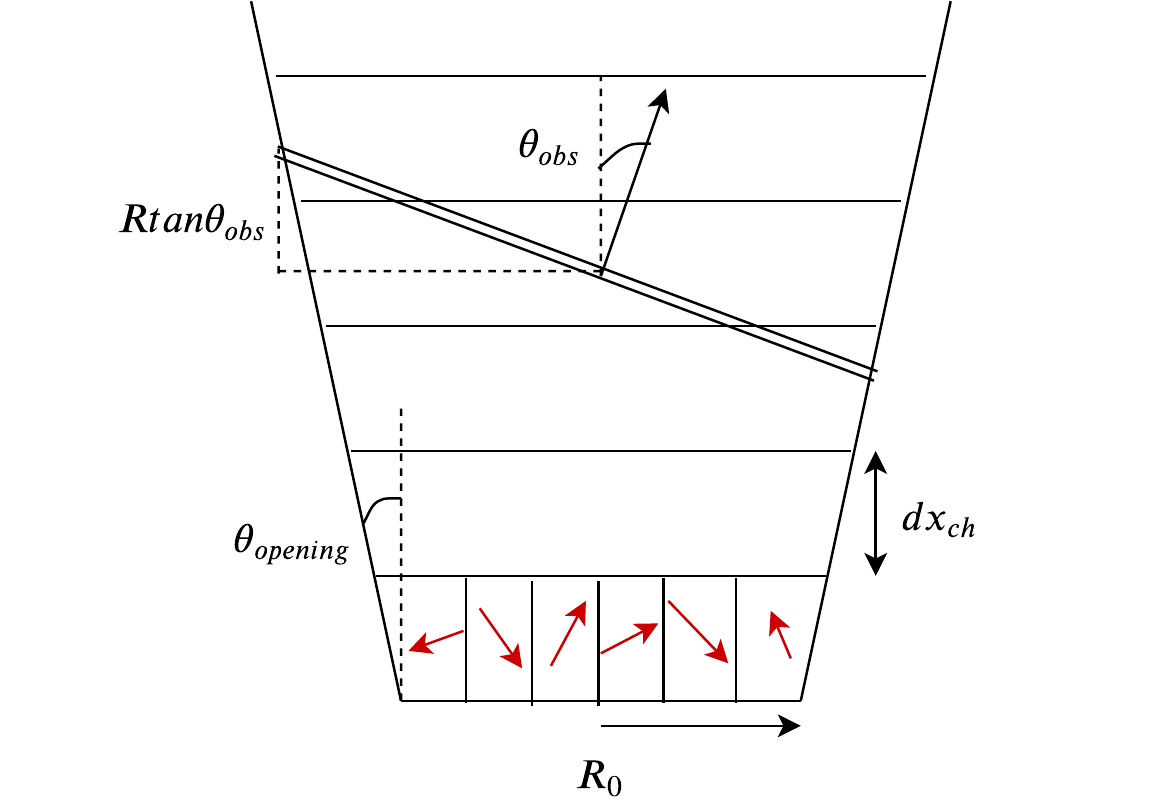}
\centering
\caption{Simplified schematic displaying the light travel effect described in \textsection 4.2. The jet is split into horizontal slices of width $dx_{\rm ch}$, the scale in the jet frame over which the magnetic field vectors decorrelate. These $B$-field vectors are denoted by the red arrows. Note that for jets viewed at large off-axis angles multiple slices are viewed at a given time.
}
\end{figure}

For typical blazar parameters $\theta_{\rm obs} \sim 0.5^{\circ} - 7^{\circ}$, $\theta_{\rm op} \sim 5^{\circ} - 45^{\circ}$, $N \sim 1 - 100$ (in a stochastic phase) and $dx_{\rm ch} \approx 2R_0/\sqrt{N}$. Then in the worst case, $2R\, {\rm tan}\,\theta_{\rm obs} / \sqrt{N} > dx_{\rm ch}$ when $R \approx 2.3R_0$. Beyond this $N_{\rm eff}$ increases while polarization fraction and variability decrease. These radii contribute most to the radio-microwave range of the synchroton peak. Through SSC this also de-polarizes the very low end of the Compton component. For rotating B-field zones the effect is similar, but slice mixing not only lowers $\Pi$ but smooths over rotational phase in the helix structure. However the large viewing angle conditions required for significant slice mixing will make the rotation less prominent, in any case.

\subsection{Finite Travel Time at Large Jet Radius: Seed Photon Build-up}

The two geometrical effects above increase $N_{\rm eff}$ slightly at large jet distance $x_p$ (affecting low energy synchrotron and SSC). But finite light travel time affects the SSC of all zones since the seed photon population in a given zone is made up of contributions from all other zones. The furthest zones are a significant light travel time away and this means that their emission represents a smaller $x$ and earlier time in the slice evolution down the jet.

The total synchrotron energy in the co-moving jet frame at any point $i$ on the jet cross section is given by:
\begin{equation}
E^i(t) = \int \frac{P^i_{\rm Sync}(r,\phi,t)}{\pi R(t)^2} rdr\,d\phi\, dz\,dt
\end{equation}
where $P^i_{\rm Sync}(x,\phi)$ is the total synchrotron power per unit length in the jet cross section. The functional form varies depending on the cross section point $i$. 
To a good approximation, the emitted synchrotron radiation from the zones is co-moving with its jet slice (\textsection4.1). This reduces the radiative transfer to a 2D sum, allowing us to set $dt = dx/c$ and $dr = dx$ where $x$ is the distance the jet cross section has travelled along the length of the jet. So the energy density is:
\begin{equation}
\begin{aligned}
\rho^i_E(x) = \frac{dE^i(x)}{dV} = \int_0^x\int_0^{2\pi} \frac{P_{\rm Sync}(x,\phi)}{2\pi^2 c R(x)^2} dx\,d\phi
\end{aligned} 
\end{equation}
For a cylindrical jet with no cooling and $x_{\rm max}>>R$, choosing $i$ to be the point in the center of the jet (16) reduces to:
\begin{equation}
\rho_E(x) =  \frac{P_{\rm Sync}}{\pi c R^2}\int_0^R x dx = \frac{P_{\rm Sync}}{\pi c R },
\end{equation}
which is a familiar expression for the energy density at the center of a 2D emitting disk.

To treat the polarization, we sum up the energy density coming from all zones in a given slice. This is done by simply evaluating the distance of every $P_{\rm Sync}(x,\phi)dx$ annulus from each of the zones in the jet cross section. Using the mutual displacement vectors between zones and their individual $\mathbf{B_i}(x)$ we can construct the total seed photon polarization and energy density at every point in the jet. For this one must compute the correct solid angle subtended by the scattering zone and the effects of RPAR rotation on the polarization vectors. Given the finite numerical nature of our simulation, we expect it to be asymptotically more accurate for a higher number of zones.

The top panel of figure 7 show this resulting seed photon population, computed using Equation (16), showing the energy density as a function of distance $x$ along the jet. The line types show the difference between edge and central zones for a conical jet. We also show how beyond a critical distance other zones in the slice dominate over self-emission in the seed photon density; this occurs later at the jet edge. Its effect can be seen on the SSC EVPA (see \textsection5). 

SSC photons of a given energy are, of course produced by a range of seed photons, so care must be taken in comparing the observed polarizations. Figure 8, shows the effective seed photon SED for X-ray (keV) and soft $\gamma$-ray (MeV) Compton emission. X-ray polarization measurements by upcoming missions are thus best compared with synchrotron observations in the mm-optical band.

\begin{figure}[]
\includegraphics[width=1.0\linewidth, height=9.2cm]{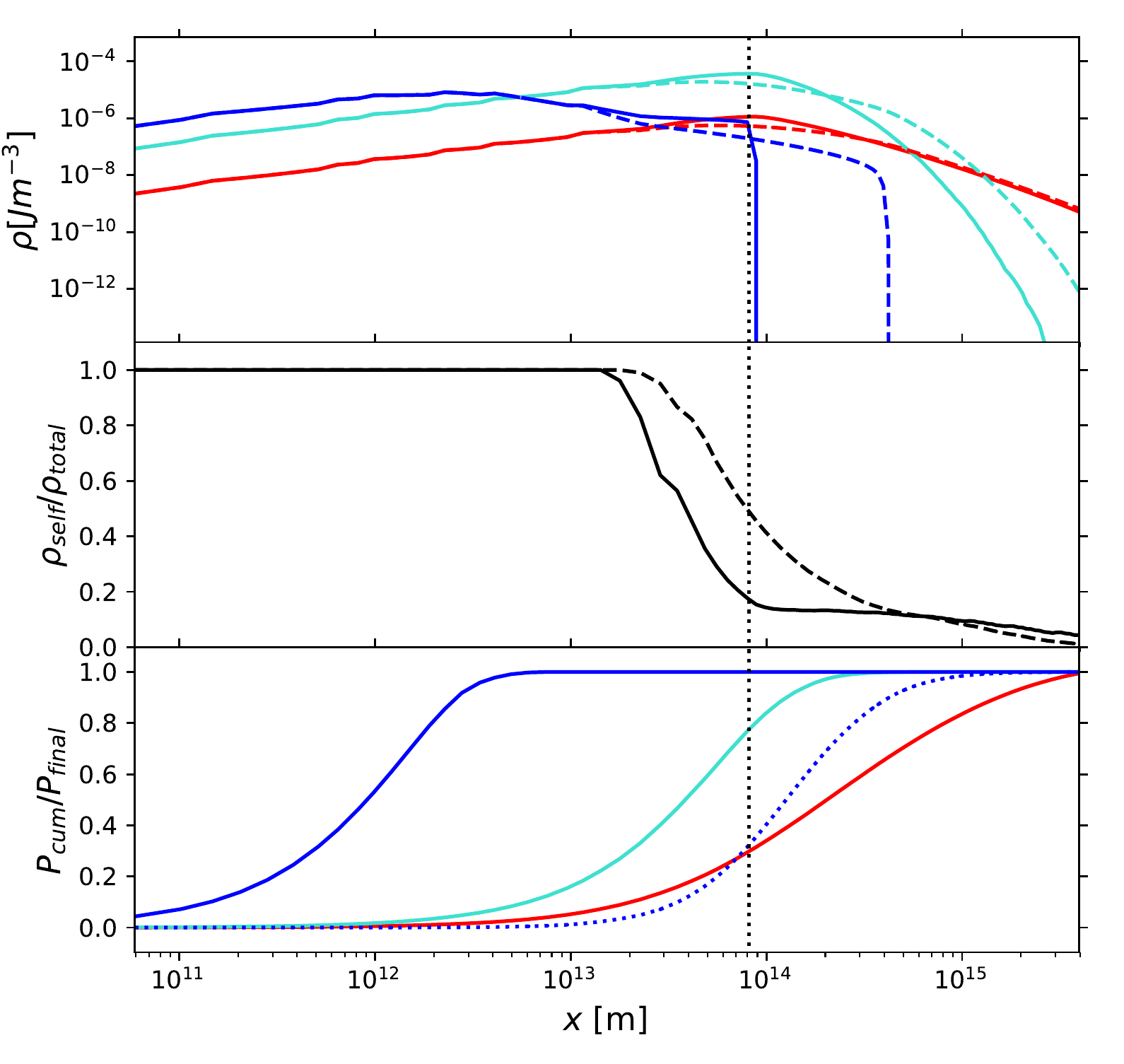}
\centering
\caption{\textit{Top panel:} Synchrotron energy density as a function of distance along the jet $x$ for three different energy bands (red: mm wavelength, cyan: optical, blue: 10 keV). Solid lines are for a the jet center, while the dashed lines are for the edge. The black vertical dotted line represent $x=R_0$. \textit{Middle panel:} Fraction of the integrated synchrotron density
due to the local zone (solid line central zone, dashed line edge zone). Neighboring zones dominate for $x > R_0$. \textit{Bottom panel:} Build up of the on-axis synchrotron seed photon population at three energies. The dotted blue line represents the corresponding 1-10\,keV SSC flux. 
These are fluxes measured in the jet frame; jet parameters are as in \S5.}
\end{figure}

\section{\label{sec:level1} SSC with All Effects}

As a concrete example, we compute with $B_i$ randomly drawn for all zones at the jet base. Orientations are frozen thereafter, e.g. fixed as the zones propagate down the jet, but magnitude can vary. As in \cite{potter_synchrotron_2012} for a ballistic conical jet we assume that the only energy loss mechanism is radiation, so magnetic energy $U_B$ is conserved. Thus magnetic flux density decreases as the jet expands. The magnetic/particle energy ratio increases slightly along the jet but remains $\sim 1$. We assume here a single fixed $\gamma_{\rm max}$ in all zones. Thus geometrical (Doppler boosting) effects dominate the prominence of individual zones. Indeed, geometric parameters ($\Gamma$, $\theta_{\rm open}$, $\theta_{\rm obs}$) have the largest effect on polarization. Other parameters ($W_j$, $\alpha$, $E_{\rm max}$, $B_0$) primarily affect the shape of the SED. As expected, polarization thus serves as an excellent (and largely independent) probe of jet geometry, although we do note when other (spectral) parameters have a large effect.

In contrast to the treatment of BCS, who assumed a simple power law electron spectrum and uniform (energy independent) synchrotron polarization, we need to consider how all seed photon energies contribute to the observed Compton radiation at a given energy in computing the ratio $\Pi_{\rm SSC}/\Pi_{\rm Sync}$. Since our electron population evolves along the jet (\S3, \S4.2), different seeds dominate at different locations along the jet. Nevertheless, we can give a qualitative picture of the seed spectrum. To connect X-ray SSC with observed synchrotron fluxes, we focus on the synchrotron seeds in the optical and mm range (see Figure 8). These are computed in simulations using all effects described above. The simulations employ various zone multiplicities (1, 7, 19, 37) to illustrate the effect the increasing the zone averaging on both the synchrotron and SSC polarization amplitudes. The principal effect is, of course a diminution $\Pi_{\rm Sync} \propto N_{\rm eff}^{-1/2}$. We compute 200+ realizations of each configuration to average down these fluctuations and display $\Pi_{\rm Sync}$, $\Pi_{\rm SSC}$ trends.

\begin{figure}[t]
\includegraphics[width=0.9\linewidth, height=5.4cm]{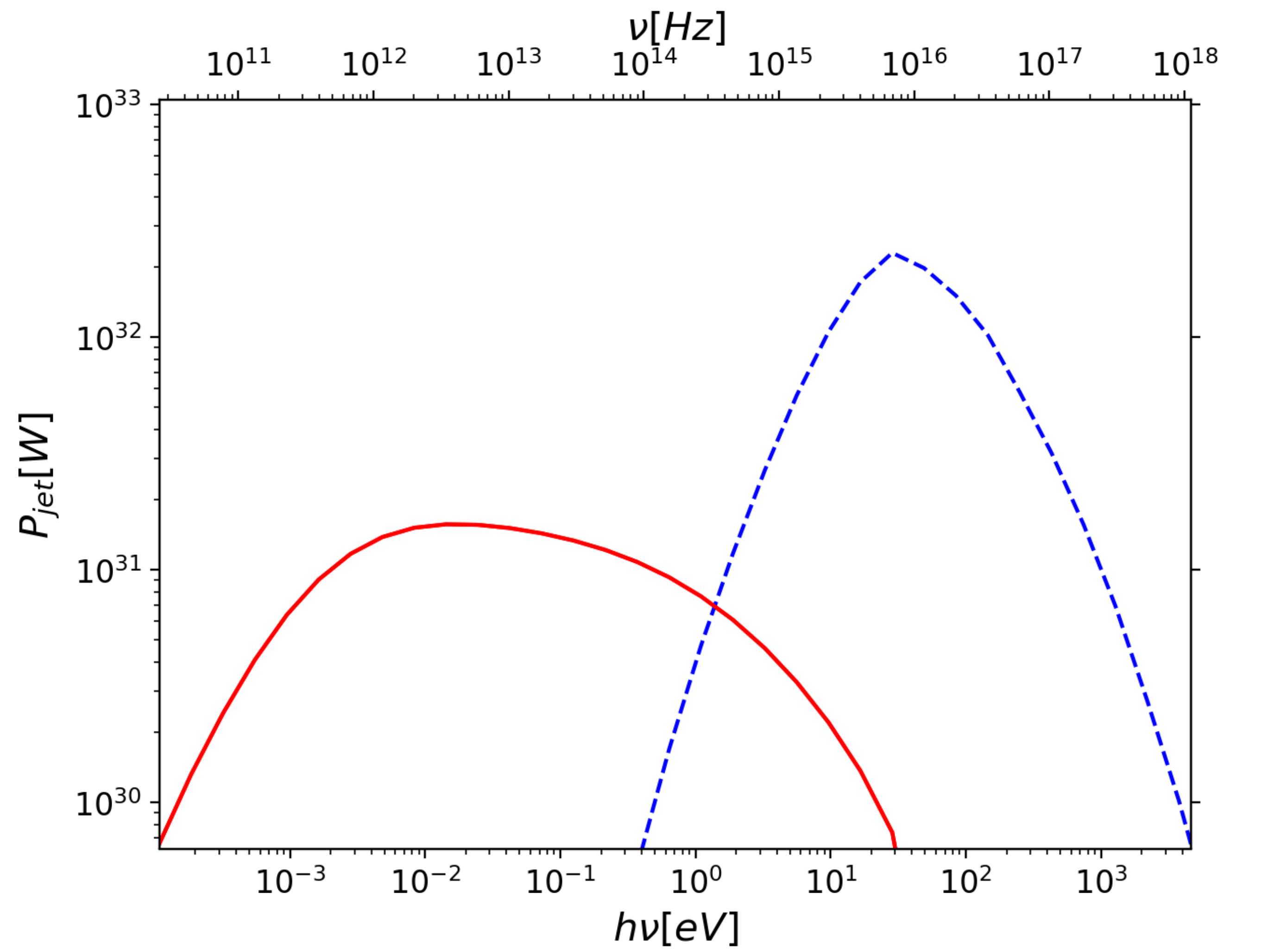}
\centering
\caption{SSC power in the jet frame as a function of the (lab frame) seed photon energy -- i.e. the seed photon spectrum for a given Compton energy. The solid line is for keV SSC emission while the dashed line shows the seed spectrum producing MeV emission. For the overall SED and assumed jet parameters, see \textsection5.1.}
\end{figure}

\begin{figure}[t]
 \includegraphics[width=0.95\linewidth, height=5.4cm]{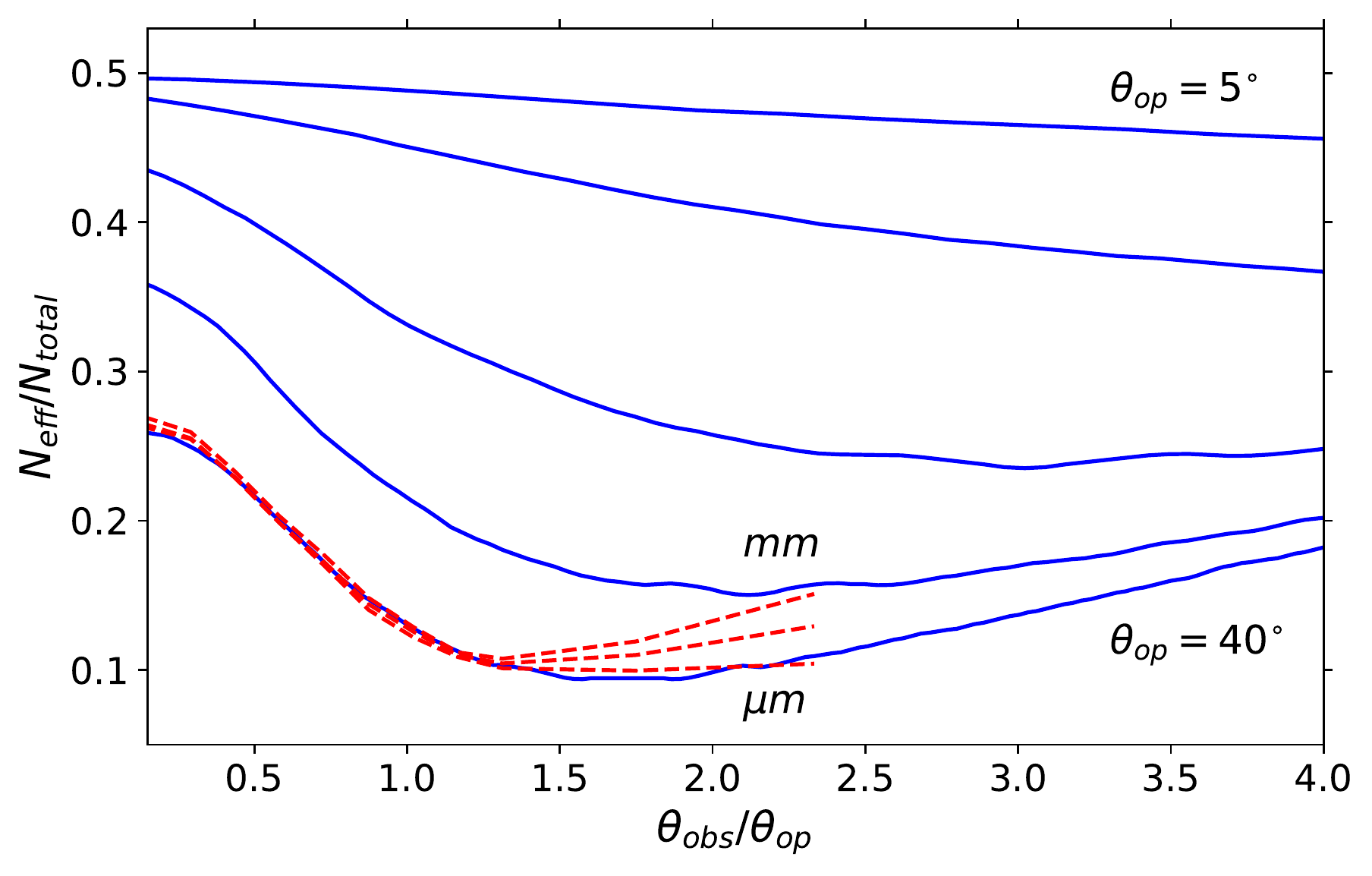}
 \centering
 \caption{The geometrical effects of Doppler boosting on the number of effective emitting zones. Jet frame $\theta_{\rm op}$ increases from top to bottom; the fraction of zones within our $1/\Gamma$ effective viewing circle decreases (blue curves). As $\theta_{\rm obs}$ increases off-axis, the fraction of the zones in the $1/\Gamma$ viewing cone also decreases, until $\theta_{\rm obs} > 1/\Gamma$ and more (distant) zones are boosted at a similar level, causing $N_{\rm eff}$ to grow again. The exact functional form is controlled by the weighting function, in this case the Doppler factor $D^4$, and the jet geometry, in this case a cone. We note that treating $N_{\rm eff}$ as the number of zones that dominates half of the flux is not a perfect proxy, the true $N_{\rm eff}$ is controlled by a weighted distribution. The red lines show the same relationship for a $\theta_{\rm op} = 40^{\circ}$ jet affected by LT effects (\textsection 4.2). At low energies (e.g. mm wavelengths), the emission occurs at sufficiently large radii (see fig.7) that light travel effects substantially increase $N_{\rm eff}$ for jets viewed off-axis. 
 }
 \end{figure}

\subsection{$N_{\rm eff}$}

In \textsection3 we discussed how the number of effective emission zones, $N_{\rm eff}$, affects the net polarization, with a large increase at synchrotron cutoff energies. $N_{\rm eff}$ effects can help explain both synchrotron and SSC polarization behavior across the whole SED. Synchrotron polarization is controlled solely by $N_{\rm eff}$, since every zone emits its synchrotron independently. Averaging over a large number of isotropic B-field iterations, we expect $\Pi_{\rm Sync} \propto N_{\rm eff}^{-1/2}$. However, the relationship between $\Pi_{\rm SSC}$ and $N_{\rm eff}$ is not {\it a priori} obvious since each single zone scatters synchrotron seed photons from all the other zones in the jet, weighted by their proximity and power. 

As in \textsection3 we define here $N_{\rm eff}$ as the minimum number of zones that contribute half of the flux. This is an imperfect estimate since $N_{\rm eff}$ depends on the weighted contribution of all zones; $\Pi_{\rm Sync}$ itself provides the best metric for $N_{\rm eff}$. In any event, the underlying behavior is adequately approximated with $N_{\rm eff} \propto \Pi_{\rm Sync}^{-2}$.

\begin{figure*}[t]
 \includegraphics[width=0.9\linewidth, height=7.6cm]{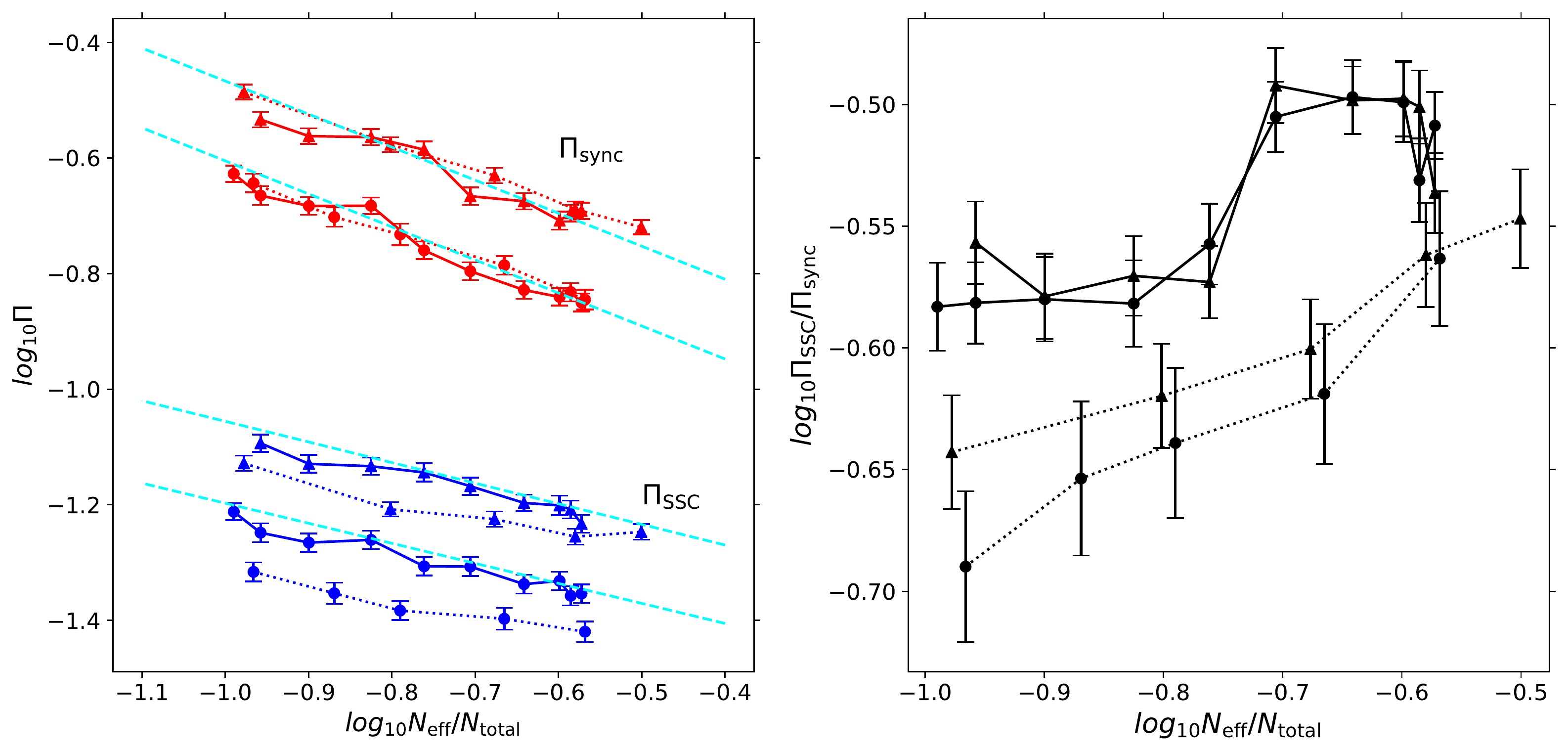}
 \centering
 \caption{\textit{Left Panel:} The optical synchrotron (red) and X-ray SSC (blue) polarization fraction plotted against $N_{\rm eff}/N$. Triangle points are for an $N=19$ zone jet, circles for $N=37$. Solid lines show a Doppler-dominated jet. For the dotted lines zone intensities are randomly distributed. The synchrotron cyan dashed lines show $N_{\rm eff}^{-1/2}$, while the SSC equivalents show the best fit power laws (N=37 gives $p=0.29$; N=19, $p=0.31$). 
 \textit{Right Panel:} 
 The SSC/synchrotron polarization ratio {\it versus} $N_{\rm eff}/N$ (line types as in left panel). Note that the SSC angle-averaging is more effective for a randomly distributed jet, leading to a lower ratio. Values are averaged over many simulations with isotropic B-fields; error bars show the residual errors on the mean. The jet parameters used in the simulations are given in the Appendix Table.}
\end{figure*}

\begin{figure}[t!]
 \includegraphics[width=0.99\linewidth, height=7.6cm]{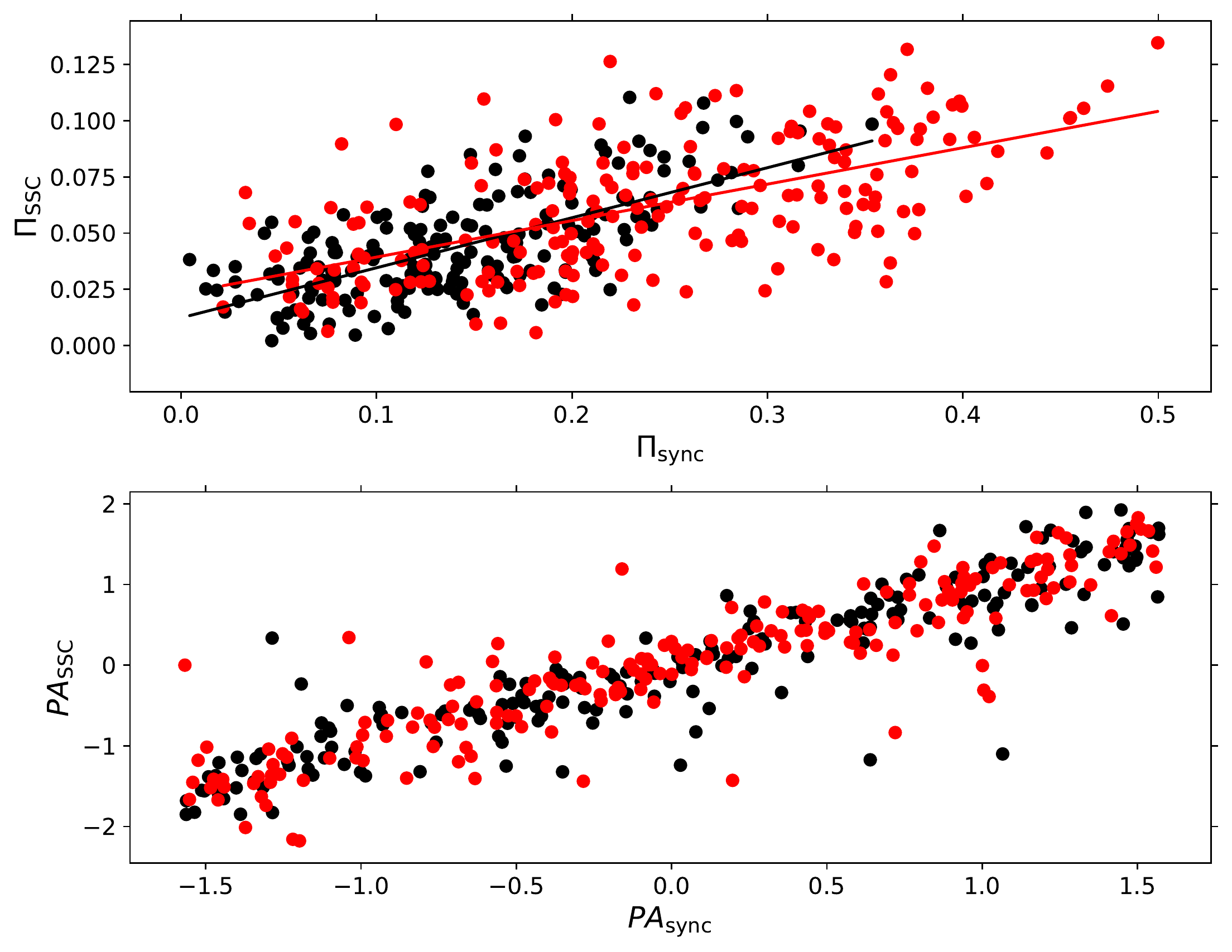}
 \centering
 \caption{Correlation plot of the X-ray SSC polarization fraction (top) and PA (bottom) against the corresponding optical synchrotron for two different viewing angles ($N_{\rm eff}$) (red: $\theta_{\rm obs} = 4.5^{\circ}$, black: $\theta_{\rm obs} = 0.1^{\circ}$). Note $\Pi_{\rm SSC}$ can persist even when chance cancellation makes $\Pi_{\rm Sync}$ for the Earth line-of-sight; this leads to high $\Pi_{\rm SSC} / \Pi_{\rm Sync} >> 1$ and a non-zero intercept.
 EVPAs are strongly correlated, but show scatter due to similar differences in the zone-sampling for the Earth line-of-sight.}
 \end{figure}
 \begin{figure}[t!]
\includegraphics[width=0.99\linewidth, height=7.6cm]{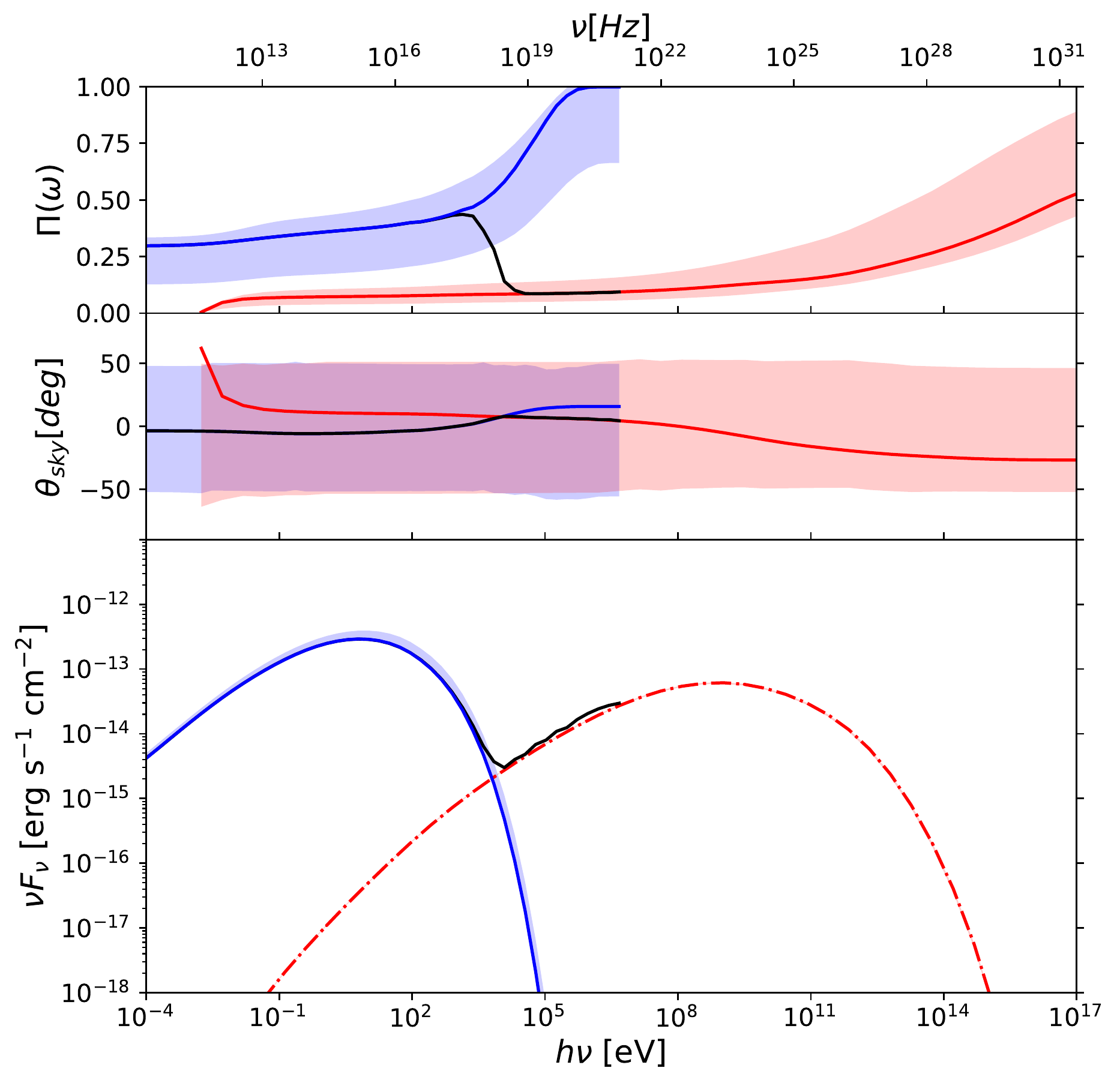}
 \centering
\caption{Example fully treated jet SED for a single random B-field draw, with $\theta_{\rm obs} = 6.0^\circ$ and the remaining jet parameters as in fig. 10. This simulation uses 19 B-field zones and serves to display some typical polarization behavior. Note the substantial jump in EVPA between the Sync and SSC components. Note also the substantial $\Pi$ increase above the synchrotron and SSC peaks due to the decreased $N_{\rm eff}$ and the average $\Pi_{\rm Sync}$ decrease for sub-infrared energies due to \textsection4.2. The shaded bands show the standard deviations in individual realization over many isotropic B-field draw; EVPA average has no preferred direction for either component, with an average value 0.}
\end{figure}

When the observed brightness of $N$ total zones is controlled purely by viewing angle-determined Doppler weighting, jet geometry determines $N_{\rm eff}/N$. In practice measuring the synchrotron polarization fraction well below the cutoff (and above radio energies affected by light travel time) provides a direct measure of $N_{\rm eff}$, and thus constrains jet geometry. Figure 9 shows the relationship between $N_{\rm eff}/N$ and $\theta_{\rm obs}/\theta_{\rm op}$ for several jet geometries (blue). For low (radio) energies (which radiate at large $x$) and large off-axis angles, finite light travel time effects in a widely diverging jet (\S4.2, Figure 6) increase $N_{\rm eff}$. Red lines in Figure 9 shows this effect (at $\theta_{\rm op} = 40^{\circ}$) for long wavelength synchrotron. 

While we expect that Doppler weighting of zone flux is always present, the observed flux from individual zones may vary for other reasons. These might include differences in acceleration efficiency between zones \citep{marscher_turbulent_2014}, with electron density, $\gamma_{\rm max}$ and $\Gamma$ variation - or variation in relative sizes of zones. Such variation may be more prominent during flaring events.
To illustrate such effects, we plot a simple model where the lab frame flux from individual zones is allowed to vary randomly (as opposed to being controlled purely by jet geometry). The results are compared with a Doppler weighted jet in figure 10 for total zone numbers $N = 19,\, 37.$

Figure 10 displays the expected $\Pi_{\rm Sync} \propto N_{\rm eff}^{-1/2}$ trend. The $N,\, N_{\rm eff}$ behavior of SSC is more complex. $N=19$ is more polarized than $N=37$ for the same $N_{\rm eff}/N$ for both synchrotron and SSC. In the synchrotron case this is simply due to the change in $N_{\rm eff}$. In the SSC case, $N_{\rm eff}$ zones are boosted and scatter the jet frame synchrotron emission from all $N$ zones. A larger $N$ further averages down the seed photon polarization (for the same $N_{\rm eff}$), so $\Pi_{\rm SSC} \propto N^{-1/2}$. This ensures that the dot point SSC curves lie below the triangle poimts in Figure 10. However $\Pi_{\rm SSC}$ also depends on $N_{\rm eff}$ (apparent in figure 10 from the non-zero slope), since with larger $N_{\rm eff}$ one has more scatterers sampling the angular distribution of the synchrotron radiation of the $N$ total zones. This averaging decreases $\Pi_{\rm SSC}$ as $N_{\rm eff}$ grows. The scaling depends on how the $N_{\rm eff}$ zones are chosen/weighted. We can characterize the dependence using a simple power law:
\begin{equation}
   \Pi_{\rm SSC} \propto N_{\rm eff}^{-p} N^{-1/2}.
\end{equation}
The Doppler-boosted cases have similar slopes $p\approx 0.3$ for both $N=19$ and $N=37$, with the decreased $\Pi_{\rm Sync}$ driving down the coefficient of the latter. Randomly selected zones (dotted lines in figure 10) give $p \approx 0.25$. However these are not universal power laws; the averaging over the seed photon's angular polarization distribution depends on the particular weighting scheme.

This can also be seen in the right panel of Figure 10. Since $p<0.5$, we have a slow increase in the SSC/Sync polarization ratio as $N_{\rm eff}$ increases, most obvious for the random zones. For pure Doppler boosting, geometrical effects complicate this trend. As $N_{\rm eff}$ grows ($\theta_{obs}$ decreases), the ratio growth is slow until one starts viewing near the jet edge (or near $1/\Gamma$). Here the increasing $N_{\rm eff}$ zones are together at the near edge of the jet; these all receive synchrotron emission from the $N$ emitting zones at similar angle. This coherence gives rise to poor averaging over the synchrotron beam and, on average, larger SSC polarization, giving an abrupt rise to the ratio. Conversely the largest $N_{\rm eff}$ occur when viewing close to the jet axis. There the most strongly boosted zones dominating the flux are nearly uniformly surrounded by their $N$ synchrotron sources, leading to better angle averaging of the synchrotron field and a drop of SSC polarization and the polarization ratio for jets viewed nearly on axis.

Other complications are also present. For example, mm wavelengths and below are emitted on average downstream from X-ray SSC (fig.7), so $\Pi_{\rm SSC} / \Pi_{\rm Sync}$ will be larger ($\sim 0.4$) for such comparison. For some jet parameters (high $\theta_{\rm obs}$, high $\theta_{\rm op}$) energies up to the optical synchrotron and X-ray SSC can also be affected by light-travel time induced $N_{\rm eff}$ increase. 

In addition to these statistical trends there is a strong correlation between $\Pi_{\rm SSC}$ and $\Pi_{\rm Sync}$ in individual realizations (Figure 11). The magnitude of the correlation depends on $N_{\rm eff}/N$: for high $N_{\rm eff}/N$ (black points) both $\Pi$ and PA are more highly correlated (small $\theta_{\rm obs}$; Spearman $r=0.7$) than the lower $N_{\rm eff}/N$ case (red points, Spearman $r=0.6$). Of course with the smaller $\theta_{\rm obs}$/larger $N_{\rm eff}$ case (black) we seldom achieve very high $\Pi_{\rm Sync}$. But when we do, we have good confidence that the SSC polarization will also be high. Notice that the intercepts are not zero; The different SSC sample can display polarization, even when the observed synchrotron polarization happens to average to near zero. This cautions us to avoid strong conclusions from one large $\Pi_{\rm SSC}/\Pi_{\rm Sync}$ measurement in a weakly polarized source, e.g. that hadronic processes are present.\\

Overall, for synchrotron seed photons emitted early in the jet (optical for the example blazar parameters) we expect to see $\Pi_{\rm SSC}/\Pi_{\rm Sync} \sim 0.3$ when $N_{\rm eff} \sim N/2$. The synchrotron and SSC polarization fractions will be strongly correlated. As $N_{\rm eff}$ decreases relative to $N$ both the ratio and correlation will decrease, controlled in detail by the zone weighting system. For mm ($\Pi$-suppressed) seed photons we expect higher ratios $\Pi_{\rm SSC}/\Pi_{\rm Sync} \sim 0.4$ but lower correlations. For the typical blazar parameters determined by the inferred opening angles (\citealp{clausen-brown_causal_2013, jorstad_kinematics_2017}), $\Gamma_{\rm bulk}$ and $\theta_{\rm obs}$ are such that $N_{\rm eff} \sim N/2$, and we expect in most cases $\Pi_{\rm SSC}/\Pi_{\rm Sync} \sim 0.3$. However for other viewing geometries the ratio can be lower. 

Since boosting is the only feasible way to change $N_{\rm eff}$ without changing $N$, we expect that non-Doppler weighting effects will preserve $N_{\rm eff} \sim N/2$ and thus $\Pi_{\rm SSC}/\Pi_{\rm Sync} \sim 0.3$.
We note that for $>$MeV energies, SSC is emitted before neighbouring zones dominate the seed photon population (fig.7). In this case the polarization amplitudes will scale as $N_{eff}^{-1/2}$ and Sync and SSC will be highly correlated, with $\Pi_{\rm SSC}/\Pi_{\rm Sync} \sim 0.35$ for the appropriate $\Pi_{\rm Sync}$ in fig. 8.
Finally, all results shown here are for electron power laws $\alpha \sim 2$. For similar systems with different power laws we expect our results for $\Pi_{\rm SSC}/\Pi_{\rm Sync}$ can be scaled with $\alpha$ as in BCS. 

\section{Conclusion}

We have shown that for a multizone relativistic conical jet, the averaging effects that control the final net polarization are sensitive to the jet opening angle and viewing geometry. In general when viewed at larger off axis angle, fewer zones contribute to the observed radiation and the residual polarization is higher. This is countered to some extent for low frequency (e.g. radio) emission, where the electrons cool slowly enough that emission comes from a large range of radii and the finite light travel time can let zones from different radii across the jet contribute at a given observation epoch -- the increase in the number of effective zones decreases $\Pi_{\rm Sync}$. Note that these trends are guaranteed by the differential Doppler effect across a conical jet, but will be obscured if electron power, $\gamma_{\rm max}$ or $\Gamma$ fluctuations dominate zone brightness variations.

One particularly interesting effect is the increased dominance of a few zones as one observes at energies well above the synchrotron peak. There the tail of the synchrotron emission is necessarily dominated by a few zones, selected either by Doppler boosting or extreme $\gamma_{\rm max}$, and $\Pi_{\rm Sync}$ increases. This also means that the EVPA converges to a direction controlled by that dominant zone, which can be quite different to that of the (lower energy) jet average. A similar effect occurs at the upper extreme of the Compton component. Thus we expect a rapid increase in polarization, and a rapid jump in EVPA, until the SSC flux overwhelms the synchrotron component, and one jumps to new SSC values (see Figure 12 for an example). This is of particular interest for `Intermediate Peak' blazars (ISP) which can have this synchro-Compton transition in the soft X-ray regime; \citet{liodakis_prospects_2019} describe this effect and its importance in selecting targets detectable to {\it IXPE} and similar X-ray polarization missions. For example, the ISP S50716+714 has an X-ray flux of $10^{-11} - 10^{-10}$erg/s/cm$^2$. Using its measured optical polarization and Fig. 5 we estimate its X-ray polarization fraction to vary between $12-30\%$. At \textit{IXPE}'s nominal sensitivity of 5.5\% MDP$_{99}$ for $10^{-11}$erg/s/cm$^2$ in 10 days, we should obtain a 99\% significance detection in $\sim 100$\,ks exposure or less. Thus variability should not strongly degrade the single epoch polarization, although longer exposures or multiple visits should see variation in $\Pi$ and EVPA. However the synchrotron emission is steeply falling in the X-ray band and detailed measurement of the polarization variation across the band may require a higher sensitivity future facility. 

For SSC polarization, the seed photons are drawn from a variety of jet zones with different $B$-field orientations. This decreases the average polarization of the seed population and hence the final Compton polarization. Since in a conical jet different jet sectors have different angles to the Earth line-of-sight and hence different boosting, the averaging is dominated by a sub-set of the jet zones and the final effects are sensitive to RPAR effects.
Nevertheless an overall trend of $\Pi_{\rm SSC} \approx 0.3 \Pi_{\rm Sync}$ (compared to optical photons) can be expected, for both Doppler zone and random zone dominated jets.

Overall, the simulations show the danger of drawing conclusions from any one realization: The scatter in $\Pi$ is comparable to $\Pi$ itself, and expected geometrical and spectral trends are only recovered when averaging over many realizations. One should also recall that external seed photons are expected to be largely unpolarized so that any EC flux will dilute the high energy polarization signal. We see that Compton polarization is understandably less powerful as a probe of jet geometry than the synchrotron signal. Nevertheless X-ray SSC polarization can be large enough to be detected in favorable cases, where comparison with the instantaneous $\Pi_{\rm Sync}$ can give (at least statistically) information on the seed fields and scattering geometry. Finally, large positive correlations between the SSC and seed synchrotron $\Pi$ make low energy polarization monitoring a useful tool for monitoring fluctuations and aiding in $\Pi_{\rm SSC}$ detection.
\bigskip

We thank I. Liodakis and A. Marsher for helpful discussions of jet polarization physics. This work was supported in part by NASA grant NNM17AA26C.

\section{Appendix}
\begin{table}[h!]
\begin{tabular}{c c c c c c c} 
\multicolumn{7}{c}{Jet Parameters}\\ \\
Fig. & $N$ & $B_0[{\rm G}]$ & $L_{\rm jet}[{\rm m}]$ & $\theta_{\rm op}[^{\circ}]$ & $\theta_{\rm obs}[^{\circ}]$ & $\Gamma$ \\ 
 \hline\hline
$2$ & $19$ & $1.0$ & $1.8 \times 10^{10}$ & $40.0$ & $1.5$ & $14.0$ \\ 
 \hline
$10$ & $19,37$ & $1.0$ & $5.3 \times 10^{15}$ & $40.0$ & $0.1-4.5$ & $14.0$ \\ 
\hline
$11$ & $37$ & $1.0$ & $5.3 \times 10^{15}$ & $40.0$ & $0.1,4.5$ & $14.0$ \\ 
\hline
$12$ & $19$ & $1.0$ & $5.3 \times 10^{15}$ & $40.0$ & $6.0$ & $14.0$ \\ 
 \hline
\end{tabular}
\caption{
Parameters used for the plotted jet models. All models additionally assume jet power $W_j = 1.3 \times 10^{37}W$, electron spectral index $\alpha = 1.85$, and electron energy range $\gamma_{\rm min}=10$ to $\gamma_{\rm max}=3.3 \times 10^{4}$.
}
\end{table}


\subsection{Computational Jet Model}
The code used for this work is made publicly available at \url{https://github.com/alpv95/SSCpol}. Further documentation on how to compile and run can be found there.

The main code consists of a C script, \textit{jet\_model.c}, that initializes a single jet slice and follows the evolution of the electron population and the emitted photon spectrum, accumulating the observed Stokes' fluxes for both synchrotron and SSC emission.  More detail on the synchrotron emission slice and its application to blazar rotations can be found in \citet{peirson_polarization_2018}. The bulk of the CPU time required is, however, spent calculating SSC emission, evaluating the integrals in Eqs. (1) and (2) $N^2 \times$ for each slice step along the jet length. This computation is accelerated using OpenMP. For example, 16 CPU cores runs the N=37 zone model (one random B-field draw, as in fig. 12) in $\sim 250$ minutes.

With the assumptions of \textsection4, each slice can be evolved independently (the exception is high observation angle mm); we do not consider synchrotron seed photons from adjacent slices. In this paper we do not include self-absorption effects which are typically significant in the longer wavelength radio emission; this has essentially no effect on the SSC X-ray fluxes. The jet is assumed to be optically thin at all times.

The algorithm begins by initializing the free jet parameters: total jet power $W_j$, bulk Lorentz factor $\Gamma_{\rm bulk}$, electron exponential energy cutoff $\gamma_{\rm max}$, observation angle $\theta_{\rm obs}$, jet opening angle (jet frame) $\theta_{\rm open}$,
electron power law index $\alpha$, initial magnetic flux density $B_0$, minimum electron energy $\gamma_{\rm min}$, number of B-field zones in jet slice $N$, the length of the jet $L_{\rm jet}$, and the number of electron energy and emitted frequency bins desired. From these, the initial jet radius $R_0$ and electron population $\frac{dN_e}{dE_e}$ discretized in energy bins can be derived following \citet{potter_synchrotron_2012}. The cross-section is split up into $N$ circular zones, with their position and mutual displacement vectors calculated. Each zone is a initialized with a B-field vector direction sampled from an isotropic distribution.

A loop over the jet length $x < L_{jet}$ begins the main calculation. The emitted synchrotron powers per unit length $P_{\perp}^i(\nu)$, $P_{||}^i(\nu)$ for each zone $i$ are calculated assuming an isotropic pitch angle distribution, following \citet{rybicki_radiative_1979}. The synchrotron photon energy density in each zone $i$ contributed by zone $j$, $\rho_{\perp}^{ij}(\nu)$, $\rho_{||}^{ij}(\nu)$, are calculated using (16). This requires keeping track of emitted synchrotron power for all prior $x$ and accounting for the RPAR between zones in the diverging jet. The integral is treated as a sum over all $x < x_{\rm current}$. The SSC power per unit length can then be calculated by treating (1) and (2) as discretized sums, resulting in $P^{\rm SSC}_{\perp}(\nu)$, $P^{\rm SSC}_{||}(\nu)$ for every zone. The electron energy losses due to emission are found, and the step length $dx$ is set by the cooling time of the highest energy occupied electron bin, with the constraint that $R_{\rm new} \leq 1.05 R$. Then $\frac{dN_e}{dE_e}$, $x$, $R$ and $B$ are updated and the emitted power for each zone is converted to a Stokes' parameter representation and boosted, using $D_i$. The loop repeats until $x \geq L_{\rm jet}$. Finally the Stokes' parameters are converted to lab frame quantities $\nu' F_{\nu'}$, $\Pi(\nu')$ and $\theta_{PA}(\nu')$ for SSC and synchrotron separately.


\bibliographystyle{apj.bst}
\bibliography{references.bib}

\end{document}